\documentclass[12pt,a4paper]{article}
\usepackage[english]{babel}
\usepackage[utf8]{inputenc}
\usepackage{amsmath}
\usepackage{amsfonts}
\usepackage{amssymb}
\usepackage{graphicx,color}
\usepackage{slashed}
\usepackage{physics}
\usepackage{tikz}
\usepackage{pgfplots}
\pgfplotsset{compat=1.17}
\usepackage{setspace}
\usepackage{parskip}
\usepackage{imakeidx}
\usepackage[margin=2.5cm]{geometry}
\usepackage[colorlinks=true, linkcolor=black, citecolor=black]{hyperref}
\usepackage{simplewick}
\usepackage{bbold}
\usepackage[hang,flushmargin]{footmisc}
\usepgfplotslibrary{polar}
\date{}
\begin{document}
\title{Quantum friction for a scalar model: spatial dependence and higher orders} 
\author{Aitor Fern\'andez and C\'esar D.~Fosco\\
{\normalsize\it Centro At\'omico Bariloche and Instituto Balseiro}\\
{\normalsize\it Comisi\'on Nacional de Energ\'{\i}a At\'omica}\\
{\normalsize\it R8402AGP S.\ C.\ de Bariloche, Argentina.} }
\maketitle
\begin{abstract} 
We use a perturbative approach to evaluate transition amplitudes corresponding to quantum friction, for a scalar model describing an atom which moves at a constant velocity, close to a material plane. In particular, we present results on the probability density per unit time of exciting degrees of freedom on specific regions of the plane. This allows one to know spatial features of the effect which could have	practical relevance, for instance, for the design of nanodevices.  

We show that the result of integrating out the probability density agrees with previous results for the same system.

We also study the effect of including higher order terms in the perturbative calculation of the probability amplitude for quantum friction, and show that they do not alter the picture obtained from the first non-trivial order, in particular, the velocity threshold for the phenomenon to occur.
\end{abstract}

\section{Introduction}\label{sec:intro} 
Fundamental particles and their interactions are inherently quantum in nature, an aspect that can occasionally lead to macroscopic manifestations in a rather straightforward way. Indeed, vacuum fluctuations have observable consequences under the appropriate circumstances; for instance, when non-trivial boundary conditions are present, as in the case of the renowned Casimir effect~\cite{libros}.  In its original realization, it involves material media imposing boundary conditions on the fluctuations of the electromagnetic (EM) field. This leads to forces or torques when the boundary conditions are time-independent, in what constitutes the static Casimir effect. One the other hand, a suitable time dependence may lead to a dissipative effect, like the creation of photons out of the vacuum, as in the dynamical Casimir effect (DCE).

In this paper,  we deal with ``non-contact friction'' or ``quantum friction'' (QF), where quantum fluctuations lead to an observable effect: a frictional force when two media move with a constant relative velocity, resulting in energy dissipation.  It is somewhat complementary to the Casimir effect, since the zero-point fluctuations of the EM field are not directly significant; rather, the EM field mediates the interaction between the fluctuating microscopic degrees of freedom on two objects. The frictional effect does not occur for perfect mirrors~\cite{Pendry97}, but it may arise in non-dispersive media~\cite{Maghrebi:2013jpa} when their relative speed exceeds the threshold set by the speed of light in the media. 

The dissipative force may also appear on a single atom moving at a constant velocity, parallel to a plate~\cite{others}, or even between two atoms moving at a constant relative velocity~\cite{barton}.  Among the various interesting of the effect, we note that there are velocity thresholds for friction to occur, related to the speed of the modes on the material media.  For instance, as shown in a recent paper~\cite{Fosco:2021aih}, for an atom near a graphene plate, the threshold is the Fermi velocity of the electrons in the plate.

In this paper, we convey results on two aspects of this phenomenon: the first is the spatial dependence of the dissipative effect; namely, which degrees of freedom on the medium are involved in the phenomenon. The second aspect we consider is that the existence and value of the velocity threshold has been, to the best of our knowledge, verified only to the first non-trivial order in perturbation theory, assuming a weak coupling between vacuum field and each media. Here we show, in a concrete model, that this feature is stable under the inclusion of higher orders in such a perturbative expansion.

These aspects are, we believe, relevant ingredients for understanding QF in the future design of quantum devices where dissipative effects may become appreciable, since they can significantly impact their efficiency and stability.  

Throughout this paper, we use a quantum field theory model which involves an atom moving at a constant speed which is parallel to a planar medium. The model we use is essentially the same as the one studied in~\cite{Farias:2019lls}, which in turn is based on~\cite{Galley:2012qz}. It consists of a vacuum real scalar field linearly coupled to a set of uncoupled quantum harmonic oscillators, which represent the microscopic ``matter'' degrees of freedom on the mirror.

Besides considering  final states where specific spatial regions on the medium are exited, we  also integrate out the transition probabilities, to provide an indirect verification of the result obtained in~\cite{Farias:2019lls}, where the total probability of vacuum decay was derived from the imaginary part of the in-out effective action to the frictional force on the plates. A similar approach has been used in~\cite{Fosco2011} and~\cite{Fosco:2007nz}, while in~\cite{BelenFarias:2014ehx} a CTP in-in formulation~\cite{CTP} was applied to calculate the frictional force between two plates in relative motion at a constant speed.

This paper is organized as follows: in Sect.~\ref{sec:thesystem}, we describe the model that we study, and also introduce some notation and conventions. In Sect.~\ref{sec:tramp}, we evaluate the transition probabilities to the lowest non-trivial order, discussing different situations, depending on the velocity $u$ of propagation of the normal modes on the medium: namely, $u > 0$ and $u=0$. For the latter, we evaluate the spatial dependence of the effect. In Sect.~\ref{sec:higher} we use the reduction formulae~\cite{Itzykson:1980rh} to analyze what happens at higher orders after noticing that a naive calculation of the next-to-leading order transition amplitude yelds a divergent result. Finally, in Sect.~\ref{sec:conc}, we present our conclusions.

\section{The system}\label{sec:thesystem}
The system we deal with here is, regarding both its dynamical variables and the interactions between them, essentially the same as the one considered in~\cite{Farias:2019lls}. It consists of a scalar variable $q(x^0)$, associated with the ``electron'': a scalar field in $0+1$ dimensions, sitting on a moving atom with a trajectory ${\mathbf r}(t)$, which is externally driven. Besides, to study friction, this trajectory will be taken to be of constant velocity, and parallel to the material plane.

The electron $q$ is coupled to a vacuum real scalar field $\varphi$ in $3+1$ dimensions, which also interacts with a medium, represented by a real scalar field in $2+1$ dimensions, $Q$, with uniform properties and concentrated on a plane.  $Q$  will have modes propagating with a velocity $u$, which we will see that is the threshold for dissipation. We shall consider in some detail the particular case $u \to 0$: {\em independent\/} oscillators and a vanishing threshold.

Regarding conventions, in this paper we shall use natural units, so that $c= 1$ and $\hbar = 1$; space-time coordinates are denoted by $x =(x^\mu)_{\mu=0}^3$, $x^0 =  t$, and we use the Minkowski metric $(g_{\mu\nu}) \equiv {\rm diag}\{ 1,-1,-1,-1 \}$. Our choice of coordinates is such that the space-time region occupied by the medium is $x^3 = 0$. Space-time coordinates relevant to the degrees of freedom on the plane shall be denoted as \mbox{$x_\shortparallel = (x^\alpha)_{\alpha=0}^2 = (t,{\mathbf x}_\shortparallel)$}. Here, ${\mathbf x}_\shortparallel \equiv (x^1,x^2)$ are two Cartesian coordinates on the spatial plane, and $\mathbf{\hat{e}}_i$ ($i = 1,\, 2,\, 3$) are the spatial basis vectors. 

With this notation, the trajectory we consider is given by:
\begin{equation}
{\mathbf r}(t) \;=\; {\mathbf v}_\shortparallel \, t \,+\, 
\mathbf{\hat{e}}_3 \, a \;,
\end{equation}
with $a$ the distance between the plane and the particle.

The total (real-time) action for the system is:
\begin{align}\label{eq:fullAction}
  \mathcal{S}[\varphi,q,Q;\mathbf{r}]=&\int\hspace{-1mm}
  d^4x\frac{1}{2}\partial_\mu\varphi\,\partial^\mu\varphi+\int\hspace{-1mm}
  d^4x~\varphi(x)J(x)+\int\hspace{-1mm} dx^0\frac{1}{2}\Big[\big(\partial_0
	q(x^0)\big)^2-\Omega_\text{e}^2 \big(q(x^0)\big)^2\Big]+\nonumber\\
    &+\int\hspace{-1mm}
    d^3x_\shortparallel\frac{1}{2}\Big[\big(\partial_0Q(x_\shortparallel)\big)^2-
	u^2\big(\nabla_\shortparallel
	Q(x_\shortparallel)\big)^2-\Omega_\text{m}^2
\big(Q(x_\shortparallel) \big)^2\Big]  \;,
\end{align}
where
\begin{equation}
 J(x)=J_\text{e}(x)+J_\text{m}(x)=g\,q(x^0)\,\delta^3\big(\mathbf{x}-\mathbf{r}(x^0)\big)+\lambda
 \,Q(x_\shortparallel)\,\delta(x^3) \;.
\end{equation}
Here, $\Omega_\text{e}$ and $\Omega_\text{m}$ are the excitation energies for the electron and the degrees of freedom of the medium respectively; and $g$ and $\lambda$ are the coupling constants.

\section{Transition amplitudes}\label{sec:tramp}
In order to study the transition amplitudes and probabilities which are
responsible for the quantum friction phenomenon, we adopt the interaction
picture, based on our choice for the free and interaction actions. In this
situation, we have the following expression for the time evolution of the 
operators corresponding to the dynamical variables:
\begin{align}
    \hat{q}(t)=& \frac{1}{\sqrt{2\Omega_\text{e}}}\left(e^{-i \Omega_\text{e}t} \hat{a} \,+\, e^{i \Omega_\text{e} t} \hat{a}^\dagger\right)\\
    \hat{Q}(x_\shortparallel)=&\int\frac{d^2\mathbf{p_\shortparallel}}{2\pi}\frac{1}{\sqrt{2p_0}}\left[e^{-ip_\shortparallel\cdot x_\shortparallel}\hat{\alpha}(\mathbf{p_\shortparallel})+e^{ip_\shortparallel\cdot x_\shortparallel}\hat{\alpha}^\dagger(\mathbf{p_\shortparallel})\right] \\
    \widehat{\varphi}(x) =&\; \int \frac{d^3{\mathbf k}}{(2 \pi)^{3/2}} \frac{1}{\sqrt{2|{\mathbf k}|}} \big( \hat{a}({\mathbf k}) \,  e^{-i k \cdot x }\,+\, \hat{a}^\dagger({\mathbf k}) \,  e^{i k \cdot x } \big) \;,
\end{align}
with $p_0=\sqrt{\Omega_\text{m}^2+u^2\mathbf{p_\shortparallel}^2}$. Note that the creation and annihilation operators do satisfy the standard
commutation relations; thus, the only non-vanishing commutators are:
\begin{equation}
[\hat{a} \,,\, \hat{a}^\dagger] \,=\, 1 \;\;,\;\;
	[\hat{\alpha}({\mathbf p_\shortparallel}) \,,\, \hat{\alpha}^\dagger({\mathbf p'_\shortparallel}) ] \,=\,
	\delta^{(2)}({\mathbf p_\shortparallel} - {\mathbf p'_\shortparallel})\;\;,\;\;
	[\hat{a}({\mathbf k}) \,,\, \hat{a}^\dagger({\mathbf k'}) ] \,=\,
	\delta^{(3)}({\mathbf k} - {\mathbf k'}) \;\;.
\end{equation} 

The transition amplitudes $T_\text{fi} = \bra{\text{f}} \widehat{T} \ket{\text{i}} $
shall be determined from the scattering matrix $\widehat{S} = \hat{I} + i
\widehat{T}$, namely, from the evolution operator in the interaction
representation,	$\widehat{U}(t_\text{f},t_\text{i})$,  for $t_\text{i} \to - \infty$ and $t_\text{f}
\to  +\infty$: 
\begin{equation}\label{eq:U}
	\widehat{S} \;=\; \widehat{U}(+\infty,-\infty) \;=\; {\rm T}  \exp \big[ i
{\mathcal S}_\text{int}(\hat{q}, \widehat{Q}, \widehat{\varphi}; {\mathbf r}(t))
\big]\;, 
\end{equation}
where ${\rm T}$ denotes the time-ordering operator and $\mathcal{S}_\text{int}$ is the interaction term of the classical action, i.e. the second term of (\ref{eq:fullAction}).

The initial quantum state $\ket{\text{i}}$ of the full system is assumed to be
the vacuum for all the modes, namely, for the electron, the medium, and the
vacuum field. In a self explanatory notation,
\begin{equation}\label{eq:in}
\ket{\text{i}};=\; |0_\text{e} \rangle \otimes |0_\text{m} \rangle \otimes |0_\text{v} \rangle \;.
\end{equation}
Regarding the final state, $\ket{\text{f}}$, in quantum friction there is no
production of vacuum-field particles (photons); indeed, that would require
a non-vanishing acceleration. Therefore, in quantum friction, only even
terms in the expansion of the exponential in (\ref{eq:U}) can intervene.
The lowest order contribution to the transition amplitude comes, therefore,
from the second order one $T^{(2)}_\text{fi}\equiv\mathcal{M}(\mathbf{p_\shortparallel})$, which yields:
\begin{equation}\label{eq:tfi}
	\mathcal{M}(\mathbf{p_\shortparallel}) \;=\; i \int d^4x \int d^4x' \, 
		(\bra{\text{f}_\text{e}} \otimes  \bra{\text{f}_\text{m}}  |) \, \widehat{J}_\text{m}(x)
		\widehat{J}_\text{e}(x') \,
		(\ket{0_\text{e}} \otimes \ket{0_\text{m}} ) \, G(x-x') \;,
\end{equation}
where we introduced the final states for the electron and the medium, and
the scalar field propagator $G$:
\begin{equation}
	G(x-x') \;=\; \int \frac{d^4k}{(2\pi)^4} \, e^{-i k \cdot (x-x')}
	\,\frac{i}{k^2 + i \varepsilon} \;,
\end{equation}
and we have assumed the initial and final states to be normalized. It is
rather straightforward to see that the only contribution to the transition
amplitude (to this order) contains a quantum for both the electron and the medium, namely:
\begin{equation}\label{eq:states}
\ket{\text{f}_\text{e}} \;=\; \hat{a}^\dagger \ket{0_\text{e}} \;\;,\;\;\;
\ket{\text{f}_\text{m}} \;=\; \frac{2\pi}{L}\hat{\alpha}^\dagger({\mathbf p_\shortparallel})
\ket{0_\text{m}} \;,
\end{equation}
where the factor $2\pi/L$ ($L$ is the length of the side a two-dimensional square box) has been introduced in order to normalize the state. The limit $L \to \infty$ at the end of the calculation is implicitly assumed. Modifications of the final state when $u\to0$ will be dealt with later.
Inserting this into (\ref{eq:tfi}), and after some operations, we obtain:
\begin{equation}
    \mathcal{M}(\mathbf{p_\shortparallel})=\frac{2\pi}{L}\frac{g\lambda}{4\sqrt{\Omega_\text{e}p_0}}\delta(p_0+\Omega_\text{e}-\mathbf{p_\shortparallel}\cdot\mathbf{v})\frac{e^{-a\sqrt{-p_\shortparallel^2}}}{\sqrt{-p_\shortparallel^2-i\epsilon}}.
\end{equation}
From the Dirac delta we see that $v\abs{\mathbf p_\shortparallel}\geq \mathbf{p_\shortparallel}\cdot\mathbf{v}=p_0+\Omega_\text{e}>u\abs{\mathbf{p_\shortparallel}}$, so in order for the transition amplitude to be non-zero, the velocity of the atom $v$ must overcome the velocity $u$. In the next section we shall see that this feature is independent on the order of approximation.

We can compute $\rho(\mathbf{p_\shortparallel})\equiv|\mathcal{M}(\mathbf{p_\shortparallel})|^2/T$, which is the probability per unit time of having the mentioned quantum state as the final state. Using
\begin{equation}
    \delta(\text{frequency})^2=\frac{T}{2\pi}\delta(\text{frequency}),
\end{equation}
we see that this quantity is, to  the lowest non trivial order in $g\lambda$,
\begin{equation}
    \rho(\mathbf{p_\shortparallel})=\left(\frac{2\pi}{L}\right)^2\frac{(g\lambda)^2}{32\pi\Omega_\text{e}}\frac{\delta\big(\sqrt{u^2\mathbf{p_\shortparallel}^2+\Omega_\text{m}^2}+\Omega_\text{e}-\mathbf{v}\cdot\mathbf{p_\shortparallel}\big)}{\sqrt{u^2\mathbf{p_\shortparallel}^2+\Omega_\text{m}^2}}\frac{e^{-2a\sqrt{(1-u^2)\mathbf{p_\shortparallel}^2-\Omega_\text{m}^2}}}{(1-u^2)\mathbf{p_\shortparallel}^2-\Omega_\text{m}^2}.
\end{equation}
Now the total probability (per unit time) is obtained by integrating the momentum, which gives
\begin{align}
    P=&\left(\frac{L}{2\pi}\right)^2\hspace{-1mm}\int\hspace{-1mm} d^2\mathbf{p_\shortparallel}~\rho(\mathbf{p_\shortparallel})\\
    =&\frac{(g\lambda)^2}{32\pi\Omega_\text{e}^3}\int d^2\mathbf{P}\frac{\delta\big(\sqrt{f^2+u^2\mathbf{P}^2}+1-\mathbf{v}\cdot\mathbf{P}\big)}{\sqrt{f^2+u^2\mathbf{P}^2}}\frac{e^{-2a\Omega_\text{e}\sqrt{(1-u^2)\mathbf{P}^2-f^2}}}{(1-u^2)\mathbf{P}^2-f^2}\label{eq:Punot0},
\end{align}
where $f=\Omega_\text{m}/\Omega_\text{e}$ and $\mathbf{P}\equiv\mathbf{p_\shortparallel}/\Omega_\text{e}$. Note that the integral is a function of adimensional quantities, $a\Omega_\text{e},\,f,\,u$ and $v$.

If we multiply the probability density by the energy transfered to the medium $E(\mathbf{p_\shortparallel})=\sqrt{u^2\mathbf{p_\shortparallel}^2+\Omega_\text{m}^2}$, we can obtain the power dissipated, and this can be related to the frictional force 
\begin{equation}
    \mathcal{W}=\left(\frac{L}{2\pi}\right)^2\int d^2\mathbf{p_\shortparallel}~E(\mathbf{p_\shortparallel})\rho(\mathbf{p_\shortparallel})=v F_\text{fr}.
\end{equation}

\subsection*{Independent oscillators ($u=0$)}
To study the case $u=0$, we also have to rethink the final state for the
medium. Now, rather than considering an excitation with fixed momentum, which now fail to exist, we should consider a localized excitation,
\begin{equation}
    \ket{\text{f}_\text{m}}=\int d^2\mathbf{x}_\shortparallel~f(\mathbf{x}_\shortparallel)\hat{\alpha}^\dagger(\mathbf{x}_\shortparallel)\ket{0_\text{m}}=\int \frac{d^2\mathbf{p_\shortparallel}}{2\pi}\tilde{f}(\mathbf{p_\shortparallel})\hat{\alpha}^\dagger(\mathbf{p_\shortparallel})\ket{0_\text{m}},
\end{equation}
where $\int d^2\mathbf{x}_\shortparallel |f(\mathbf{x}_\shortparallel)|^2=1$. The second equallity 
helps us to accommodate this to the description used previously
\begin{equation}
    \mathcal{M}_{u=0}[f]=\int\frac{d^2\mathbf{p_\shortparallel}}{2\pi}\left[\tilde{f}(\mathbf{p_\shortparallel})\right]^*\frac{L}{2\pi}\mathcal{M}(\mathbf{p_\shortparallel})|_{u=0}=\frac{L}{2\pi}\int\frac{d^2\mathbf{p_\shortparallel}}{2\pi}\tilde{f^*}(\mathbf{p_\shortparallel})\mathcal{M}(-\mathbf{p_\shortparallel})|_{u=0}.\label{eq:Mu0}
\end{equation}
The $L/2\pi$ factor has been introduced because $\mathcal{M}(\mathbf{p_\shortparallel})$ was computed with the final state $\frac{2\pi}{L}\hat{\alpha}^\dagger(\mathbf{p_\shortparallel})\ket{0_\text{m}}$, but now we want to use the transition amplitude corresponding to $\hat{\alpha}^\dagger(\mathbf{p_\shortparallel})\ket{0_\text{m}}$.

For the expression (\ref{eq:Mu0}) we get
\begin{equation}
    \mathcal{M}_{u=0}[f]=\frac{ig\lambda}{8\pi\sqrt{\Omega_\text{e}\Omega_\text{m}}}\int d^2\mathbf{p_\shortparallel}\,\tilde{f^*}(\mathbf{p_\shortparallel})\delta(\Omega_\text{m}+\Omega_\text{e}+\mathbf{v}\cdot\mathbf{p_\shortparallel})\frac{e^{-a\sqrt{\mathbf{p_\shortparallel}^2-\Omega_\text{m}^2}}}{\sqrt{\mathbf{p_\shortparallel}^2-\Omega_\text{m}^2}}
\end{equation}

Choosing $f(\mathbf{x}_\shortparallel)=L~\delta^{(2)}(\mathbf{x}_\shortparallel-\mbox{\boldmath$ \xi$})$,\footnote{The $L$ factor is needed in order for the final state to be normalized. We have used 
$$\delta^{(2)}(\mbox{\boldmath$ \xi$}-\mbox{\boldmath$ \xi$})=\int\limits_{-\pi/L}^{\pi/L}\frac{d^2\mathbf{p_\shortparallel}}{(2\pi)^2}=\frac{1}{L^2}$$} the probability per unit area of finding the excited oscillator located at $\mbox{\boldmath$ \xi$}$ is given by  
\begin{equation}\label{eq:probDens}
    \frac{1}{L^2}\abs{\mathcal{M}_{u=0}[\delta_\xi]}^2=\frac{(g\lambda)^2}{(8\pi)^2 v^2 \Omega_\text{e}\Omega_\text{m}}\abs{\int dp_\perp~e^{-ip_\perp\xi_\perp}\frac{e^{-a\sqrt{p_\perp^2-\Omega^2}}}{\sqrt{p_\perp^2-\Omega^2}}}^2,
\end{equation}
 where $\Omega^2=(\Omega_\text{m}+\Omega_\text{e})^2/v^2-\Omega_\text{m}^2$, and in this expression $\shortparallel$ and $\perp$ mean respectively parallel and perpendicular to the velocity of the atom. 

\begin{figure}
    \centering
    \begin{tikzpicture}
        \begin{axis}[
            width=9cm,height=7.2cm,
            xmin=0,
            xmax=0.1,
            ymin=-0.1,
            ymax=+2.5,
            xlabel=$v$,
            ylabel=$P$,
            y label style={rotate=270},
            xtick={0.0,0.02,0.1},
            extra tick style={grid=major},
            legend style={at={(axis cs:0.01,1.5)},anchor=south west},
            legend cell align=left,
            y tick label style={
            /pgf/number format/.cd,
            fixed,
            precision=2,
            /tikz/.cd}
        ]
        \addplot+ [
            smooth,
            blue,
            tension = 0.7,
            mark=none,
            ultra thick,
        ]
        file {probu.dat};
        \addlegendentry{$u=0.02$},
        
        \addplot+ [
            smooth,
            orange,
            tension = 0.7,
            mark=none,
            ultra thick,
        ]
        file {probu0.dat};
        \addlegendentry{$u=0$},
        \end{axis}
    \end{tikzpicture}
    \caption{Dependence of total probability with the velocity of the atom. For both curves $\Omega_\text{e}a=0.01$ and $f=0.5$.}
    \label{fig:probability}
\end{figure}
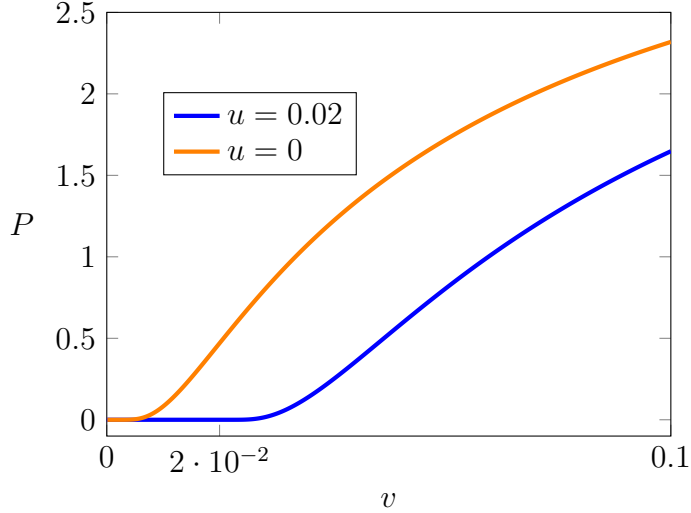

 To compute the total probability, we have to integrate (\ref{eq:probDens}) over the whole plane. Actually, since this quantity only depends on $\xi_\perp$, we have to multiply by a characteristic length in the parallel direction instead of integrating over it. This characteristic length can be taken as $vT$, and we can think of $T$ being the time that the atom is moving. Then we can divide this quantity by $T$ to get the probability per unit time, and it turns out to be
\begin{equation}\label{eq:probu0}
    P=\frac{1}{T}vT\int d\xi_\perp\frac{1}{L^2}\abs{\mathcal{M}_{u=0}[\delta_\xi]}^2=\frac{(g\lambda)^2}{32\pi\Omega_e^3}~\frac{1}{vf}\int dP_\perp\frac{e^{-2a\Omega_\text{e}\sqrt{P_\perp^2+g^2}}}{P_\perp^2+g^2},
\end{equation}
that coincides with the expression (69) of \cite{Farias:2019lls}, where the same probability was calculated using the in-out effective action formalism\footnote{Recall that $P=2\Im\Gamma$}. Note that we could have obtained the same result by taking $u=0$ in (\ref{eq:Punot0}). We show in Figure \ref{fig:probability} how this probability changes with the velocity of the atom, for both cases when $u\neq0$ (previous subsection) and $u=0$.

From (\ref{eq:probu0}) we see that $v\abs{\mathcal{M}_{u=0}[\delta_\xi]}^2/L^2\equiv\rho(\xi_\perp)$ is the probability density (per unit time) of finding the excited oscillator located at a perpendicular coordinate $\xi_\perp$, which is
\begin{equation}
    \rho(\xi_\perp)=\frac{(g\lambda)^2}{(4\pi)^2\Omega_\text{e}^2}~\frac{1}{v f}\abs{\int\limits_0^\infty dP_\perp\cos(P_\perp~\Omega_\text{e}\xi_\perp)\frac{e^{-a\Omega_\text{e}\sqrt{P_\perp^2+g^2}}}{\sqrt{P_\perp^2+g^2}}}^2,
\end{equation}
with $P_\perp=p_\perp/\Omega_\text{e}$ and $g^2=(1+f)^2/v^2-f^2$. This function gives us information about the spatial dependence of quantum friction. 

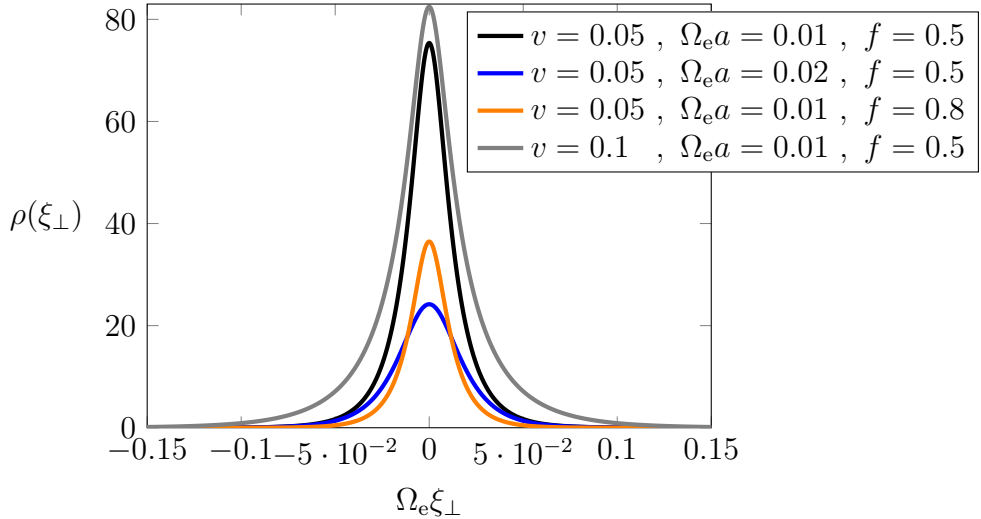
\begin{figure}
    \centering
    \begin{tikzpicture}
        \begin{axis}[
            width=9cm,height=7.2cm,
            xmin=-0.15,
            xmax=0.15,
            ymin=0,
            ymax=+83,
            xlabel=$\Omega_\text{e}\xi_\perp$,
            ylabel=$\rho(\xi_\perp)$,
            y label style={rotate=270},
            xtick={-0.15,-0.1,-0.05,0.0,0.05,0.1,0.15},
            extra tick style={grid=major},
            legend style={at={(axis cs:0.02,50.0)},anchor=south west},
            legend cell align=left,
            y tick label style={
            /pgf/number format/.cd,
            fixed,
            precision=2,
            /tikz/.cd}
        ]
        \addplot+ [
            smooth,
            black,
            tension = 0.7,
            mark=none,
            ultra thick,
        ]
        file {v005f05a001.dat};
        \addlegendentry{$v=0.05~,~\Omega_\text{e} a=0.01~,~f=0.5$},
        
        \addplot+ [
            smooth,
            blue,
            tension = 0.7,
            mark=none,
            ultra thick,
        ]
        file {v005f05a002.dat};
        \addlegendentry{$v=0.05~,~\Omega_\text{e} a=0.02~,~f=0.5$},
        
        \addplot+ [
            smooth,
            orange,
            tension = 0.7,
            mark=none,
            ultra thick,
        ]
        file {v005f08a001.dat};
        \addlegendentry{$v=0.05~,~\Omega_\text{e} a=0.01~,~f=0.8$},
        
        \addplot+ [
            smooth,
            gray,
            tension = 0.7,
            mark=none,
            ultra thick,
        ]
        file {v01f05a001.dat};
        \addlegendentry{$v=0.1~~\,,~\Omega_\text{e} a=0.01~,~f=0.5$},
        \end{axis}
    \end{tikzpicture}
    \caption{Probability density for different values of $v$, $a$ and $f$, with fixed $\Omega_\text{e}$.}
    \label{fig:spatialDependence}
\end{figure}
In Figure \ref{fig:spatialDependence} we can see how this distribution depends on the parameters of the system. First, we see that the distribution is centered and has its maximum at $\xi_\perp=0$, i.e. the region immediately below the atom. When the distance between the atom and the medium increases (blue vs. black), the probability decreases everywhere, but the width of the distribution does not change that much. This width is related to the scope of the effective interaction between the atom and the medium. On the other hand, when the quotient between the frequencies of excitation increases (orange vs. black), not only the probability decreases, but also the width of the curve does it  considerably. Finally, we see that when the velocity of the atom increases (gray vs. black), the most noticeable effect is that the scope of the effective interaction increases.

In this case, the energy transferred to the medium is constant, so the dissipated power is just
\begin{equation}
    \mathcal{W}=\Omega_\text{m}P=vF_\text{fr}.
\end{equation}

\section{Higher orders}\label{sec:higher}

Let us evaluate here the effect of including higher orders in the calculation of the transition amplitudes, for the same kind of of quantum friction process we have considered in the previous Section.  As we shall see, this requires a calculation which takes into account a renormalization of the parameters in action for the model. In other words, a standard evaluation of the subsequent terms in the pertubative expansion of the transition amplitude, does not produce meaningful results. This is the case when computing the next-to-leading order contribution, which turns out to be divergent due to the fact that the electron correlator appears evaluated right at the value where it has a pole.

When this happens, a particularly convenient way to obtain the transition amplitudes, is to use the reduction formulae~\cite{Itzykson:1980rh}, adapted to the case at hand.  Indeed, that approach allows to obtain the amplitudes from the corresponding Green's functions: vacuum expectation values of the relevant time-ordered products.  

On the other hand, in the QF processes we deal with here, the initial state $\ket{\text{i}}$ of the full system is assumed to be the vacuum for all the degrees of freedom, namely, for the electron, the medium, and the vacuum field.  Regarding the final state, $\ket{\text{f}}$, it will also be the vacuum for the scalar field. 
Thus, the Green's functions we need  have no external legs of the real scalar field $\varphi$. This implies, in a path integral approach (that we shall adopt), the real scalar field may be integrated out in order to evaluate those Green's functions. 
To that end, we introduce $\Gamma[q,Q;{\mathbf r}]$, a sort of ``intermediate effective action'', defined by:
\begin{equation}\label{eq:defgamma}
	e^{i \Gamma[q,Q;{\mathbf r}]}\;=\;\frac{\int {\mathcal D}\varphi \; 
e^{ i \mathcal{S}[\varphi,q,Q;\mathbf{r}]}}{\int {\mathcal D}\varphi \; 
e^{ i {\mathcal S}_0[\varphi]}} \;, \;\;\;\;
{\mathcal S}_0[\varphi]\;\equiv\; \int d^4x
\frac{1}{2}\partial_\mu\varphi\,\partial^\mu\varphi \;.
\end{equation}
Thus,
\begin{equation}
 \Gamma[q,Q]=S_\text{e}[q]+ S_\text{m}[Q]+\frac{i}{2}\int d^4x\int
 d^4y\,J(x)G(x-y)J(y) \,.
\end{equation}
where $S_\text{e}[q]$ and $S_\text{m}[Q]$ are the free actions for the atom
and the medium (third and fourth terms of (\ref{eq:fullAction})), and
$G(x-y)$ is the free Feynman propagator of the massless scalar field
\begin{equation}
G(x-y)=\int\frac{d^4 k}{(2\pi)^4}\frac{i}{k^2+i\epsilon}e^{-ik\cdot
(x-y)}\;.
\end{equation}
Therefore, the relevant Green's functions can be obtained as functional
integral correlation functions of $q$ and $Q$, evaluated with $\Gamma$
playing the role of action, namely:
\begin{equation}
\langle q(x^0) q(y^0) \ldots Q(x_\shortparallel) Q(y_\shortparallel)\ldots
\rangle \;=\; \frac{\int {\mathcal D}q \int {\mathcal D}Q \, q(x^0) q(y^0)
\ldots Q(x_\shortparallel) Q(y_\shortparallel)\ldots \
e^{i \Gamma[q,Q;{\mathbf r}]}}{\int {\mathcal D}q
\int {\mathcal D}Q \, e^{i \Gamma[q,Q;{\mathbf r}]}}\;.
\end{equation}

Besides, the {\em connected\/} correlation functions, due to the structure
of $\Gamma$, have two arguments, namely: $\langle Q(x_\shortparallel)
Q(y_\shortparallel)\rangle$, $\langle
q(x^0) q(y^0)\rangle$, and $\langle q(x^0) Q(y_\shortparallel)\rangle$. 
Those involving the same variable are just the full propagators and will
only account for a renormalization of the model's parameters. Therefore,
we are led to consider $\langle q(x^0) Q(y_\shortparallel)\rangle$, and
from this (via reduction formulae) we shall extract then the only
non-trivial friction amplitude. Recalling the fact that the initial state
is assumed to be the vacuum for all the fields, the only possibility of
having then a non-vanishing transition amplitude is to have, as final
states, both the atom and the medium in an excited state. In order to write
explicitly the transition amplitudes, we introduce:
\begin{align}
\Delta_{qq}(x^0,y^0) \,\equiv \,\langle q(x^0) q(y^0) \rangle
\;\;&,\;\; 
\Delta_{QQ}(x_\shortparallel,y_\shortparallel) \,\equiv\,\langle
Q(x_\shortparallel) Q(y_\shortparallel) \rangle \nonumber\\  
\Delta_{qQ}(x^0,y_\shortparallel) \,\equiv \,\langle q(x^0)
Q(y_\shortparallel) \rangle \;\;&,\;\;
\Delta_{Qq}(x_\shortparallel, y^0) \,\equiv\,\langle
Q(x_\shortparallel) q(y^0) \rangle \;.
\end{align}
These, in turn, may be obtained by first noting that $\Gamma$ has the
structure of a quadratic (functional) form in $q(x^0)$ and
$Q(x_\shortparallel)$ which, using a shorthand  notation
for the integrals~\footnote{For example, $\int_{x^0,y_\shortparallel}
\ldots = \int dy^0 \int d^3y_\shortparallel \ldots $, etc.}, is given explicitly by:
\begin{align}
	\Gamma[q,Q] &= \frac{i}{2}\Big[ \int_{x^0,y^0} 
	q(x^0)K_{qq}(x^0,\,y^0) q(y^0) 
	\,+\,\int_{x^0,\,y_\shortparallel}  q(x^0)K_{qQ}(x^0,y_\shortparallel)
	Q(y_\shortparallel)  \nonumber\\
	&+\,\int_{x_\shortparallel,\,y^0}  Q(x_\shortparallel)
	K_{Qq}(x_\shortparallel,y^0) q(y^0) \,+\,
	\int_{x_\shortparallel,\,y_\shortparallel} Q(x_\shortparallel)
	K_{QQ}(x_\shortparallel,y_\shortparallel) Q(y_\shortparallel) 
	\Big] \;,
\end{align}
where:
\begin{align}
K_{qq}(x^0,y^0)\,=& i(\partial_{x^0}^2+
\Omega_\text{e}^2)\,\delta(x^0-y^0)\,+\,g^2
G\big(x^0-y^0,\mathbf{r}(x^0)-\mathbf{r}(y^0)\big) \\
 K_{QQ}(x_\shortparallel,y_\shortparallel)\,=&\,
i(\partial_{x^0}^2-u^2\mathbf{\nabla}_\shortparallel^2+\Omega_\text{m}^2)\,\delta^3(x_\shortparallel-y_\shortparallel)+\lambda^2G(x_\shortparallel-y_\shortparallel)\\
  K_{qQ}(x^0,y_\shortparallel)=&\,g\lambda\,
G\big(x^0-y^0,\mathbf{r}(x^0)-\mathbf{y}_\shortparallel\big)\\
 K_{Qq}(x_\shortparallel,y^0)=&\,\lambda
	g\,G\big(x^0-y^0,\mathbf{x}_\shortparallel-\mathbf{r}(y^0)\big) \;.
\end{align}
The correlation functions are then obtained from the inverse of $[K]$, a
$2\times 2$ matrix of kernels,
\begin{equation}
	[K]\;\equiv\; 
\left( 
	\begin{array}{cc}
		K_{qq} & K_{qQ} \\
		K_{Qq} & K_{QQ}
	\end{array}
\right) \;,
\end{equation}
in terms of its inverse $[K]^{-1}$, as follows:
\begin{equation}\label{eq:expr}
 i \; [K]^{-1}\;=\; 
\left( 
	\begin{array}{cc}
	\Delta_{qq} & \Delta_{qQ} \\
        \Delta_{Qq} & \Delta_{QQ}
	\end{array}
\right) \;.
\end{equation}
Namely,
\begin{align}
\Delta_{qq}(x^0,y^0) \,=\,i K^{-1}_{qq}(x^0 , y^0) 
\;\;&,\;\; 
\Delta_{qQ}(x^0,y_\shortparallel) \,=\, i K^{-1}_{qQ}(x^0,y_\shortparallel) 
\nonumber\\  
\Delta_{Qq}(x_\shortparallel,y^0) \,=\,i K^{-1}_{Qq}(x_\shortparallel, y^0) 
\;\;&,\;\; 
\Delta_{QQ}(x_\shortparallel,y_\shortparallel) \,=\, i
K^{-1}_{QQ}(x_\shortparallel,y_\shortparallel) \;.
\end{align}	
It is perhaps worth remarking that the elements of the inverse of $[K]$
appearing above are not the inverses of the matrix elements of $[K]$
(except when $g = 0$ or $\lambda = 0$).
Rather, (\ref{eq:expr}) implies
the relations:
\begin{align}\label{eq:relations}
\int_{z^0} K_{qq}(x^0,z^0) \Delta_{qq}(z^0,y^0) + \int_{z_\shortparallel}
K_{qQ}(x^0,z_\shortparallel) \Delta_{Qq}(z_\shortparallel,y^0) 
& = i \delta(x^0-y^0) \nonumber\\
\int_{z^0} K_{qq}(x^0,z^0) \Delta_{qQ}(z^0,y_\shortparallel) + 
\int_{z_\shortparallel} K_{qQ}(x^0,z_\shortparallel)
\Delta_{QQ}(z_\shortparallel,y_\shortparallel) & = 0 \nonumber\\
\int_{z^0} K_{Qq}(x_\shortparallel,z^0) \Delta_{qq}(z^0,y^0) + 
\int_{z_\shortparallel} K_{QQ}(x_\shortparallel,z_\shortparallel)
\Delta_{Qq}(z_\shortparallel,y^0) & = 0 \nonumber\\
\int_{z^0} K_{Qq}(x_\shortparallel,z^0) \Delta_{qQ}(z^0,y_\shortparallel) 
+ \int_{z_\shortparallel} K_{QQ}(x_\shortparallel,z_\shortparallel) 
\Delta_{QQ}(z_\shortparallel,y_\shortparallel) & = i
\delta^2(x_\shortparallel-y_\shortparallel)\;.
\end{align}
Thus, in general, with $A, B$ taking the values $q, Q$: $K^{-1}_{AB} \neq
(K_{AB})^{-1}$.

From (\ref{eq:relations}), we get the exact relations 
\begin{align}
	K^{-1}_{qQ}(x^0,y_\shortparallel) &=\; - \int_{z^0,
	w_\shortparallel} \, K^{-1}_{qq}(x^0,z^0)\,
	K_{qQ}(z^0, w_\shortparallel) \,
	(K_{QQ})^{-1}(w_\shortparallel,y^0) \nonumber\\
	[K^{-1}_{QQ}]^{-1}(x_\shortparallel,y_\shortparallel) &=\; 
	 K_{QQ}(x_\shortparallel,y_\shortparallel) \,-\, 
	\int_{z^0, w^0} \, K_{Qq}(x_ \shortparallel,z^0)\,
	(K_{qq})^{-1}(z^0,w^0) K_{qQ}(w^0, y_\shortparallel) \,
\end{align}
Therefore, in a shorthand notation whereby we omit the integrals, we get:
\begin{equation}
(\Delta_{qq})^{-1} \, \Delta_{qQ} \, (\Delta_{QQ})^{-1} \,=\, - i \,\Big[
K_{qQ} - K_{qQ} (K_{QQ})^{-1} K_{Qq} (K_{qq})^{-1} K_{qQ}  \Big] \;.
\end{equation}

To proceed, we recall that the initial ($\ket{\text{i}}$) and final
($\ket{\text{f}}$) states contain, by assumption, no quanta of the vacuum
field $\varphi$, while the states of the relevant degrees of freedom, $Q$
and $q$ are best introduced from the
expansion in creation and annihilation operators of the free solutions:
\begin{align}
\hat{Q}(x_\shortparallel)=&\int\frac{d^2\mathbf{p}_\shortparallel}{2\pi}\frac{1}{\sqrt{2p_0}}\left[e^{-ip_\shortparallel\cdot
x_\shortparallel}\hat{\alpha}(\mathbf{p}_\shortparallel)+e^{ip_\shortparallel\cdot
x_\shortparallel}\hat{\alpha}^\dagger(\mathbf{p}_\shortparallel)\right]\\
\hat{q}(y^0)=& \frac{1}{\sqrt{2\Omega_\text{e}}}\left(e^{-i \Omega_\text{e}
	y^0} \hat{a} \,+\, e^{i \Omega_\text{e} y^0} \hat{a}^\dagger
\right) \;,
\end{align}
with $p_0=\sqrt{\Omega_\text{m}^2+u^2\mathbf{p}_\shortparallel^2}$. 
The only non-vanishing commutation relations are:
\begin{equation}
[\hat{\alpha}(\mathbf{p}_\shortparallel)\,,\, 
\hat{\alpha}^\dagger(\mathbf{p'}_\shortparallel)]\,=\, 
\delta^2(\mathbf{p}_\shortparallel - \mathbf{p'}_\shortparallel) \;\;,\;\;\;
[\hat{a}\,,\, \hat{a}^\dagger ] \,=\; 1 \;.
\end{equation}
Thus, we consider: 
\begin{equation}
\ket{\text{i}}\,=\,\ket{0_\text{m}} \otimes \ket{0_\text{e}}
\;\;,\;\; \ket{\text{f}}\,=\,
\ket{\text{f}_\text{m}(\mathbf{p}_\shortparallel)}
\otimes\ket{\text{f}_\text{e}} \;,
\end{equation}
with $\ket{\text{f}_\text{m}(\mathbf{p}_\shortparallel)}=
\frac{2\pi}{L}\hat{\alpha}^\dagger(\mathbf{p}_\shortparallel)\ket{0_\text{m}}$,
and $\ket{\text{f}_\text{e}}=\hat{a}^\dagger \ket{\text{0}_\text{e}}$,
where the factor $\frac{2\pi}{L}$ ($L$ is the length of the side a two-dimensional
square box) has been introduced in order to normalize the state. The limit
$L \to \infty$ at the end of the calculation is implicitly assumed.

Finally, reduction formulae yield the transition amplitude $T_{fi}$ by
contracting with the inverses of the full propagators of the external
lines, and attaching the properly normalized wave functions:
\begin{equation}\label{eq:texact}
T_{fi} \,=\, - \frac{1}{2 L} \int dx^0 \int d^3y_\shortparallel 
\frac{e^{ i q^0 x^0 + i p_\shortparallel \cdot y_\shortparallel}}{\sqrt{p^0({\mathbf
p}_\shortparallel) q^0}}\, 
\big[ K_{qQ}(x^0,y_\shortparallel) \,-\, L_{qQ}(x^0, y_\shortparallel)\big]
\;, 
\end{equation}
where $q^0$ and $p_\shortparallel$ are assumed to be on-shell; to the
lowest order, that corresponds to:  $q^0 =
\Omega_\text{e}$ and $p^0 = \sqrt{ u^2 {\mathbf p}_\shortparallel^2 +
\Omega_\text{m}^2}$, and we have introduced: 
\begin{equation}
L_{qQ}(x^0, y_\shortparallel) \,\equiv\, \;\big[ K_{qQ} (K_{QQ})^{-1}
K_{Qq} (K_{qq})^{-1} K_{qQ}\big](x^0, y_\shortparallel) \;.
\end{equation}

The leading contribution comes from the first term in (\ref{eq:texact}), and noting that 
\begin{align}\label{eq:ftK}
\int dx^0 \int d^3y_\shortparallel \, e^{i q^0 x^0 + i
p_\shortparallel \cdot y}\, K_{qQ}(x^0,y_\shortparallel) &=\,
	\widetilde{K}_{qQ}(q^0,-p_\shortparallel) \nonumber\\
= - \, 2 \pi i \, \delta( p^0 + q^0 - {\mathbf
	p}_\shortparallel \cdot {\mathbf v})  \,& \, \lambda g \, 
\frac{e^{- a \sqrt{-p_\shortparallel^ 2 - i \epsilon}}}{2 \sqrt{-p_\shortparallel^ 2 
- i \epsilon}}
\;,
\end{align}
where the tilde signals Fourier transformation, we get the same result as in Section \ref{sec:tramp}. 

In order to evaluate the correction to the leading order
term, we also need to Fourier transform the other kernels appearing in the exact formula for
$T_{fi}$. These are:
\begin{align}
\tilde{K}_{qq}(p^0,q^0) \;=\; & -2 \pi i\; \delta(p_0-q_0) \, 
\left[p_0^2-\Omega_\text{e}^2 - \Pi_\text{e}(p_0)\right] \nonumber\\
\tilde{K}_{QQ}(p_\shortparallel,q_\shortparallel)\;= \; & -(2\pi)^3i
\;\delta^3(p_\shortparallel-q_\shortparallel) \,
\left[p_0^2-u^2\mathbf{p}_\shortparallel^2 - \Omega_\text{m}^2 -
\Pi_\text{m}(p_\shortparallel) \right] \nonumber\\
\tilde{K}_{Qq}(p_\shortparallel,q^0)\; =\; & - 2\pi i \; 
\,\delta\big(p_0-q_0-\mathbf{p}_\shortparallel
	\cdot\mathbf{v}\big) 
\lambda g \, \frac{e^{- a \sqrt{-p_\shortparallel^ 2 - i \epsilon}}}{2
\sqrt{-p_\shortparallel^2 
- i \epsilon}} \;
\end{align}
with:
\begin{align}
\Pi_\text{e}(p_0)=& - ig^2 \,
\int\frac{d^3\mathbf{k}}{(2\pi)^3}\tilde{G}(p_0+\mathbf{k}_\shortparallel\cdot\mathbf{v},
	\mathbf{k})\\
\Pi_\text{m}(p_\shortparallel)=& - i\lambda^2
\,\int\frac{dk^3}{2\pi}\tilde{G}(p_\shortparallel,k^3) \;.
\end{align}
Using these expressions, we can Fourier transform $L_{qQ}(x^0,y_\shortparallel)$, and we obtain that it is also proportional to the same Dirac delta of (\ref{eq:ftK}), which means that the threshold for the quantum friction to happen is the same for all orders.

It is worth pointing out that  $i (K_{qq})^{-1}$ and $i (K_{QQ})^{-1}$, which
appear in $L_{qQ}$ are not the correlators for $q$ and $Q$, respectively.
In fact, their form could be found from the exact equations:
\begin{align}
\big(\Delta_{qq} \big)^{-1} &=\; - i [ K_{qq} - K_{qQ} (K_{QQ})^{-1} K_{Qq} ]
\nonumber\\
\big(\Delta_{QQ}\big)^{-1}  &=\; - i [ K_{QQ} - K_{Qq} (K_{qq})^{-1} K_{qQ}
	] \;,
\end{align}
and these show that the corresponding self-energies are rather involved
and depend on all the matrix elements of $K$, and therefore are non trivial
functions of the distance $a$.



\section{Conclusions}\label{sec:conc}
In this paper, we have evaluated the transition amplitudes corresponding to the elementary processes which  lead to the phenomenon of quantum friction between a moving atom and a material plane. Our choice of system and model, allows for a determination of the spatial dependence of the Casimir friction phenomenon, by providing the functional form of the amplitude as a function of the distance, on the plane, to the projection of the atom's trajectory. The dependence of the previous results on the distance between the atom and the plane is modulated by a precise combination of the model's parameters and the velocity of the atom. 

By integrating out the probabilities to this order, we have found agreement with the total vacuum decay probability obtained, for the same model, by an evaluation of the imaginary part of the effective action, in a functional integral approach.

We have also shown the consequences of having a nonzero velocity for the excitations of the material plane, of which the most remarkable one is the appearance of a threshold for the velocity of the atom. 

Finally, after exploring the exact expression for the transition amplitude by means of the reduction formula, we have obtained that the threshold is the same for every order in the coupling constants.

\section*{Acknowledgements}
The authors thank ANPCyT, CONICET, CNEA and UNCuyo for financial support.

\end{document}